# SHAPE – A SPECTRO-POLARIMETER ONBOARD PROPULSION MODULE OF CHANDRAYAAN-3 MISSION

Anuj Nandi, Swapnil Singh, Bhavesh Jaiswal, Anand Jain, Smrati Verma, Reenu Palawat, Ravishankar B. T., Brajpal Singh, Anurag Tyagi, Priyanka Das, Supratik Bose, Supriya Verma, Waghmare Rahul Gautam, Yogesh Prasad K. R., Bijoy Raha, Bhavesh Mendhekar, Sathyanaryana Raju K., Srinivasa Rao Kondapi V., Sumit Kumar, Mukund Kumar Thakur, Vinti Bhatia, Nidhi Sharma, Govinda Rao Yenni, Neeraj Kumar Satya, Venkata Raghavendra, Vivechana M. S., Evangelin Leeja Justin, Praloy Karmakar, Anurag Patra, Naga Manjusha J., Motamarri Srikanth, Chinmay Kumar Rajhans, Kalpana K., Veeramuthuvel P.

Scientist/Engineer
U. R. Rao Satellite Centre (URSC)
Indian Space Research Organisation (ISRO), Department of Space (DOS)
Bangalore – 560017, India
Email: anuj@ursc.gov.in

## Abstract

***SHAPE** (**S**pectro-polarimetry of **HA**bitable **P**lanet **E**arth) is an experiment onboard Chandrayaan-3 Mission for the study of spectro-polarimetric signatures of the habitable planet Earth in the near-infrared (NIR) wavelength range (1.0 – 1.7 µm). The spectro-polarimeter is the only scientific payload (experimental in nature) on the Propulsion Module (PM) of Chandrayaan-3 mission. The instrument is a compact and light-weight spectro-polarimeter with an Acousto-Optic Tunable Filter (AOTF) at its heart. The AOTF operates in the frequency range of 80 MHz to 135 MHz with a power of 0.5 – 2.0 Watts. The two output beams (e-beam and o-beam) from the AOTF are focused onto two InGaAs detectors (pixelated, 1D linear array) with the help of focusing optics. The primary (aperture) optics of ~ 2 mm diameter collects the NIR-light for input to the AOTF, defining the FOV of 2.6°. The payload is realized with a mass of 4.8 kg and a power of 25 Watts. In this manuscript, we highlight some of the ground-based results including the post-launch initial performance of the payload while orbiting around the Moon to observe Earth.*

## Introduction

The Chandrayaan-3 (Ch-3) mission was designed and configured to carry the Lander and Rover Modules with five payloads. However, the Propulsion Module (PM) which has carried the Lander and Rover modules was initially configured without any scientific payloads. Later, as an opportunity, SHAPE, an experimental payload was accommodated in the propulsion module. The accommodation and mounting configuration along with FOV clearance of SHAPE payload on PM is shown in Fig. 1.

The payload is mounted on the –PITCH panel of PM with view direction along the +ROLL axis. The instrument is a compact and light-weight spectro-polarimeter with an Acousto-Optic Tunable Filter (AOTF) at its heart. The AOTF consists of a birefringent crystal (TeO$_2$). Based on the principle of Bragg's diffraction inside the crystal, it acts as a tunable filter, tuned by an external RF source (Goutzoulis and Pape, 1994; [1]). The primary telescope collects the NIR-light for input to the AOTF. On the application of RF signal, the AOTF splits the light into 3 beams – one unpolarised broadband beam and two orthogonally polarised narrow band beams - the Extra-ordinary (e) and Ordinary (o) beams. The unpolarised beam is absorbed using a beam absorber and the two polarised beams from the AOTF are focused onto two Indium Gallium Arsenide (InGaAs) detectors (pixelated, 1D linear array: Detector-1 and Detector-2) with the help of focusing optics. The concept design of the SHAPE payload including all the major components is shown in Fig. 2.

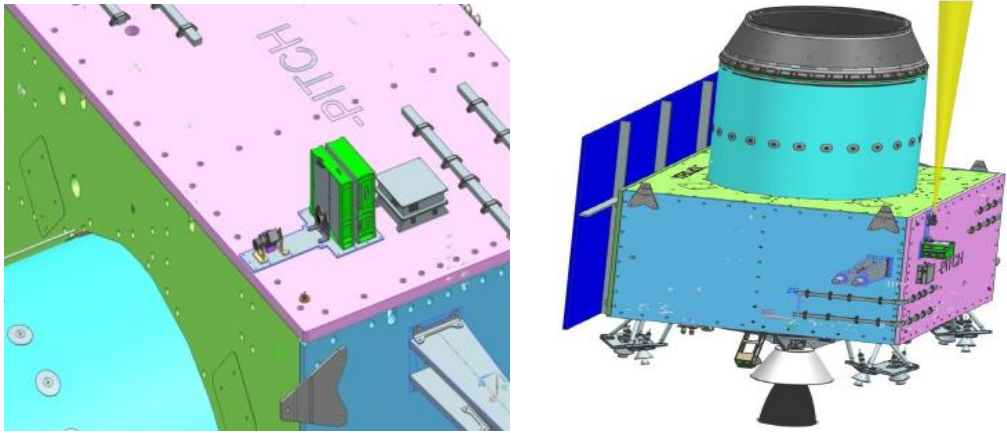

*Fig. 1: Mounting configuration of SHAPE payload on Propulsion Module with clearance of FOV.*

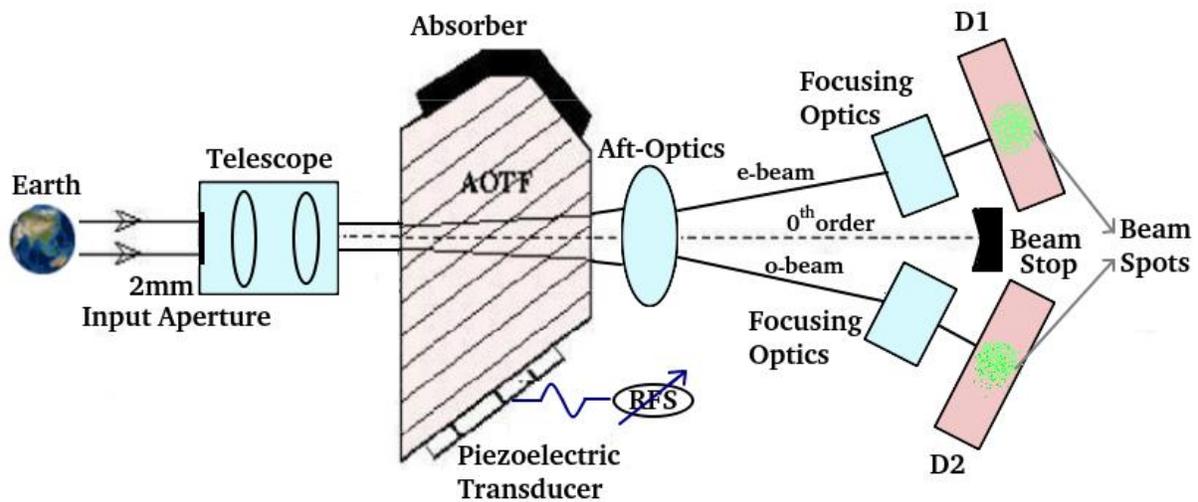

*Fig. 2: Concept design of SHAPE payload. All major components (Telescope, AOTF, Detectors, RFS, Aft-Optics, Focusing-Optics) are also marked.*

At ISRO, the development of AOTF-based IR spectrometer for space explorations was initiated a decade back (Nandi et al. 2013; [2]). Prior to the development of SHAPE payload, the spectro-polarimetric capabilities of AOTF have been demonstrated in the laboratory (see Agrawal et al. 2015; [3] and Jaiswal et al. 2022; [4] for details) of Space Astronomy Group at URSC. Due to lesser weight and compact nature, AOTF based spectrometers have been flown in various space missions (Korablev et al. 2017; [5]).

## Science Objectives

SHAPE is designed to perform spectro-polarimetric observations of Earth at various phase angles and to study the temporal changes in these observations. The major science objectives of SHAPE include temporal, spectral and polarimetric studies, which are mentioned below,

**Temporal Studies:**

The disc-integrated flux from Earth shows a periodic variation, which is correlated with the rotation of Earth as various continents and oceans move in and out of the FOV. Each orbit of Chandrayaan-3 is 2 hours (approx.), which allows SHAPE to perform Earth observations once every two hours. Hence, a light curve can be generated by combining the various observations within a day, which would allow us to infer the existence of water oceans and continents along with temporally small-scale variations such as cloud cover in a day.

**Spectral Studies**:

SHAPE will observe the spectrum in the 1.0 to 1.7 μm wavelength range in which the spectral fingerprints of various gases (e.g. $H_2O$, $CO_2$, $O_2$) are expected to be visible. The NIR spectrum is also sensitive to ice and water cloud reflectance and this study can be directly applied to study the cloud cover for characterizing a habitable 'exoplanet', where liquid water is expected.

**Polarimetric Studies:**

SHAPE would carry out 'first of its kind' disc-integrated observation of polarization of Earth. Polarization is caused by the scattering from the clouds with some contribution from sharp reflections from reflecting surfaces. SHAPE observations will aid in segregating the clouds based on altitude due to the dependence of polarization signatures on the type of clouds. It is also expected to study the phenomenon of 'glint' (direct reflection of sunlight from oceans).

## Design Configuration

SHAPE payload is designed with three packages with following details,

1. Electro-Optical Detector System (EODS)-Optics: EODS-Optics package housed with Optics, Opto-mechanical assembly, AOTF, Detector and Front-end Electronics to process the detector signals.

2. EODS-Electronics: This package houses the Processing Electronics (PE) and Power Distribution Electronics (PDE) along with two DC-DC converters.

3. Radio Frequency Source (RFS): The RFS sub-system housed with RF Synthesizer, Driver Amplifier, Power Amplifier and Interface Electronics along with one DC-DC converter to generate RF power (0.5 – 2.0 Watts) in the frequency range of 80 to 135 MHz to drive the AOTF for optimal performance.

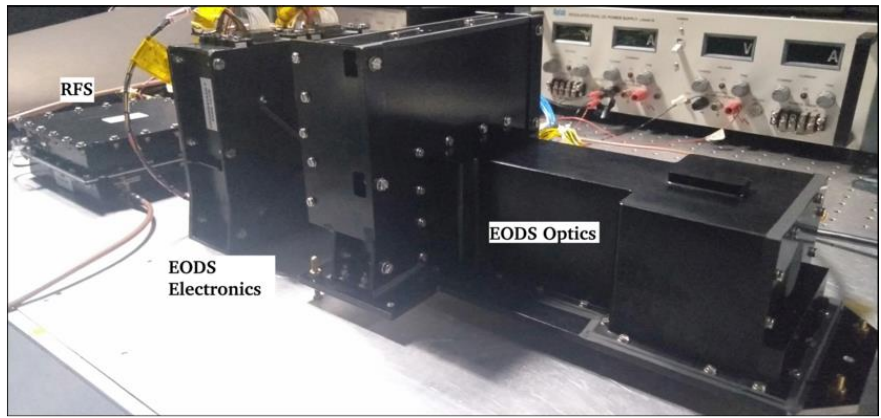

*Fig.3: Flight configuration of SHAPE payload. All three packages (EODS-Optics, EODS-Electronics and RFS) are also marked.*

The EODS-Optics package is specially designed with thermal control system (Bhatia et al. 2023; [6]) as required for optimal performance of detectors, AOTF and optical elements. The end-to-end design of SHAPE payload is presented in Configuration Design Review (CDR) document (Nandi et al. 2021; [7]). The flight configuration of the payload is shown in Fig. 3.

SHAPE is a very versatile payload, where multiple parameters can be commanded as required for the scientific observations. The RF range and the RF step sizes can be altered by command including two special values of RF step sizes of 10 kHz and 20 kHz for calibration activities. The full RF range scan is from 80 to 135 MHz including capabilities to carry out sub-scans by commanding the start, stop and step frequencies. As the payload carries out observations for different phase angles of Earth, the flux from Earth will vary with phase angles. To cater to the flux variation, the instrument has been configured with seven different integration times of 10, 20, 50, 100, 200, 500 and 1000 ms. The instrument has two modes of data storage: (i) Individual Pixel Mode (IPM) and (ii) Sum Mode (SM). In the IPM, the data for the pixels are written individually, while in the SM the data for the pixels are summed up and then stored. From the Lunar orbit, the subtended angle of the Earth's disc is around 2° in size and the instrument FOV is designed for 2.6°. For a disc of 2°, the spot size on the pixelated detectors falls on 4 pixels. The number of pixels for which the data is stored, is also commandable with a maximum of 10 pixels incorporating the movement of Earth in the FOV. The major specifications of the SHAPE payload are tabulated in Table 1.

**Table-1: Specifications of the payload**

| Specification | Parameter/Value |
|---|---|
| Spectral Range | 1.0 – 1.7 μm |
| Spectral Resolution | 2 – 4 nm |
| RF Range | 80 – 135 MHz |
| RF Power | 0.5 – 2.0 Watts |
| RF Step | 0.2 – 2.0 MHz |
| Detector | InGaAs (pixilated, 1D linear array) |
| Detector Noise | < 2.66 mV |
| FOV | 2.6° |
| Data Volume | 17 kBytes |
| Power | 24.8 Watt |
| Mass | 4.8 kg |

## Ground Test & Calibration

The SHAPE payload was extensively tested on ground. Firstly, each of the subsystems was individually tested to ensure optimal performance. The flight detectors were characterized for their pixel-wise performance. Comprehensive testing of the EODS electronics along with detectors was carried out to minimize the noise of the instrument for the operating range of temperatures (-3 $^0$C to +3 $^0$C). A noise level of < 2.66 mV was achieved with the flight design of EODS electronics. Fig. 4 shows the noise performance for one of the pixels in each detector. Along with noise, the detectors were also characterized for the variation in dark signal with integration time. Interface test was also carried out to test the variation of RF output power with frequency and the AOTF response with varying frequency and power.

After the subsystem level tests, all the subsystems were integrated and further tests were carried out at payload level. These tests (Singh et al. 2022; [8]) included the Initial Bench Test (IBT), Vibration Test, Thermo-vacuum Test, Thermal Balance Test, Electromagnetic Induction and Electromagnetic Conduction Tests, and the Final Bench Test (FBT) before delivery of the payload for integration with spacecraft.

During each of these tests both IPM and SM of the payload were tested and the key parameters such as the noise and the spectral resolution were monitored.

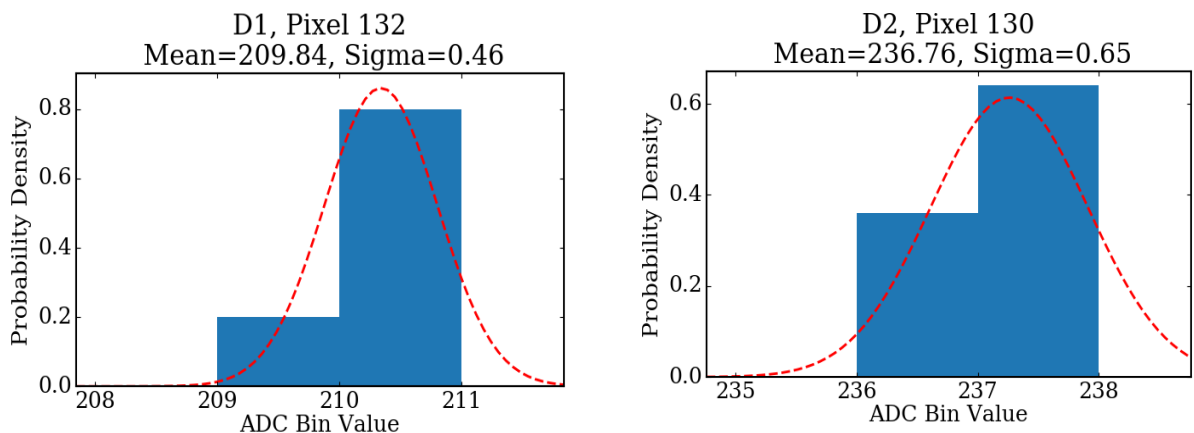

*Fig. 4: Histogram for one pixel in each detector. The dashed red line is the Gaussian fit to the histogram, where the mean value gives the dark signal in the pixel and the sigma value is used to estimate the noise.*

Detailed on-ground calibration of the payload was carried out after the FBT. The ground calibration includes radiometric, spectral, polarimetric, and field calibration. The radiometric calibration is carried out to convert the obtained ADC channel values to intensity units. The spectral response function of the AOTF is a *sinc$^2$* function. The spectral response function was measured across the entire wavelength ($\lambda$) range at steps of 50 nm and the FWHM of the function was used to estimate the spectral resolution. The spectral response of AOTF at 1400 nm is shown in Fig. 5. On application of RF, the AOTF filters out a particular wavelength and the relation between the input RF and output wavelength is called the RF-$\lambda$ relation. It is measured using the standard halogen lamp emission lines and scanning across the frequency range. The RF range of 80-135 MHz corresponds to a wavelength range of 1.7-1.0 µm and the spectral resolution varies from 4 - 2 nm in this wavelength range. The polarimetric calibration includes the study of the rotation modulation of the instrument, when a linearly polarized light was input into it.

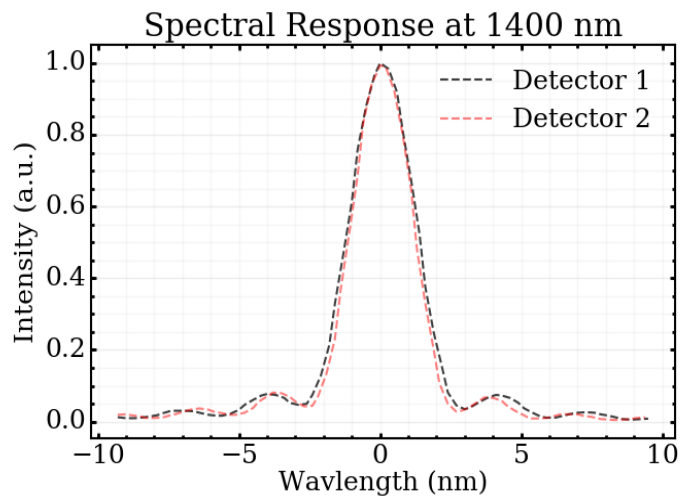

*Fig. 5: The measured spectral response of the SHAPE payload at 1400 nm for both detectors.*

As per design, the instrument has an FOV of 2.6° and the field calibration is carried out to study the variation across this FOV. To carry out this calibration, a flat source is input to the instrument at different input angles and the instrument response is measured. The field variation for both detectors is shown in Fig. 6.

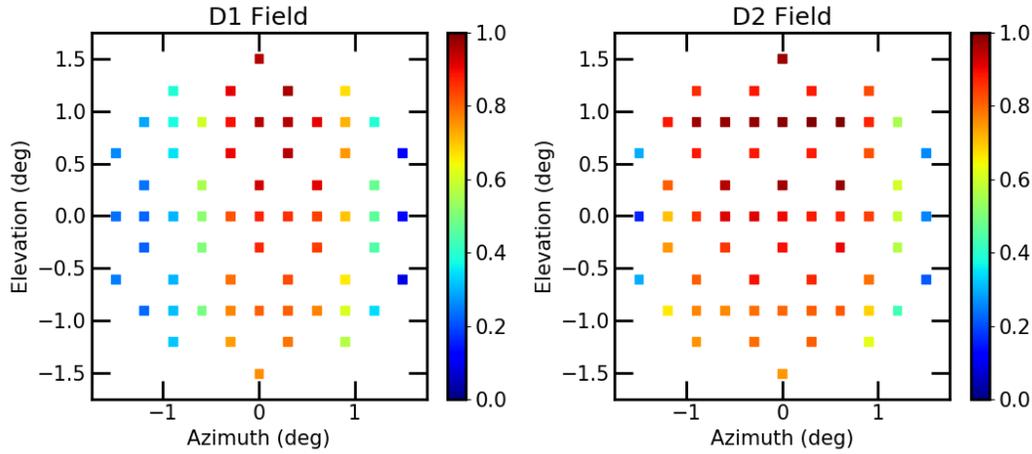

*Fig. 6: Field calibration measurements for both detectors at 1400 nm.*

Detailed ground calibration aspects along with on-board spectral and polarimetric calibration will be presented elsewhere (Jaiswal et al. 2024; [9]). The results from the calibration have been incorporated in a Calibration Database (CalDB), which will be input into the data pipeline to generate the science products.

After integration of the payload with the spacecraft, dis-assembled and assembled mode tests were carried out to verify the payload performance. Fig. 7 shows the Krypton lamp spectra recorded in the SM of the instrument during the satellite integrated tests. Auto-compatibility tests were also carried out to test the influence of all the other spacecraft components on the payload performance and vice versa. After successful completion of all tests, the Propulsion Module along with the Lander Module was transported to SHAR to integrate with the launch vehicle.

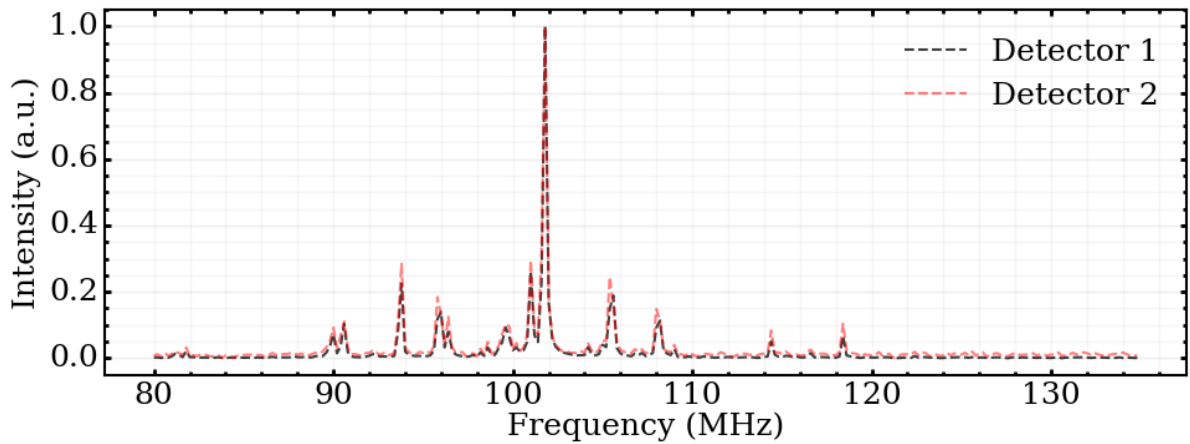

*Fig. 7: Krypton lamp spectra recorded by SHAPE payload for both detectors during the dis-assembled mode test after integration with the Propulsion Module.*

# Early In-flight Operation of SHAPE

In every ~ 2 hrs orbital time, the SHAPE operations last for typically a little over 30 min, during which the Ch-3/PM would be oriented to bring the SHAPE's optical axis aligned along the Earth vector. In the absence of SSR and BDH on Ch-3/PM, SHAPE has been designed with a data interface entirely on the Telemetry (TM) chain. Also, this set-up leads to the primary operational constraint of Ground Station visibility for scheduling observations with SHAPE in order to ensure Real-Time dump over TM and Programmable Telemetry (PTM). The Aux data (consisting of attitude Qs and latched OBT samples from the payload and OBC) are designed to use PTM Format-4 (Nandi et al. 2021; [7]), which gets dumped during the SHAPE observations. In comparison, the payload data has been designed to get stored in the payload's FPGA during observations and is dumped separately via playback after the observations are completed. The Thermal constraints may not allow to keep the payload ON during the entire orbit and therefore it is essential to command playback soon after observation before the payload is switched off. The maximum amount of memory that can be stored in the payload's FPGA is 17 kBytes comprising of 8 kBytes data per detector and 1 kByte of HK data. The PM only has a Thruster for attitude control with a dead-band of $0.75^O$ during the Earth observation. For calibration, SHAPE is frequently scheduled to observe the Moon as well. The integration time during these observations is to be set as per the phase angle.

SHAPE was commissioned on August 21, 2023, during the orbit number 152 between 08:33 – 09:07 UT. The very first pointing was towards the Earth, for the complete RF band of 80 – 135 MHz, at a RF-step size of 200 kHz with the AOTF power set to 0.5 W, and for an integration time of 200 ms. Subsequently, in the next ten days or so, gradually the different operating parameters were optimized. The AOTF power was varied for 1.0 W as well as 1.5 W, the start pixel and the number of pixels read were adjusted for the optimum beam illumination, and different integration times of 100 ms, 50 ms and 10 ms were set for different observations to regularize the observations.

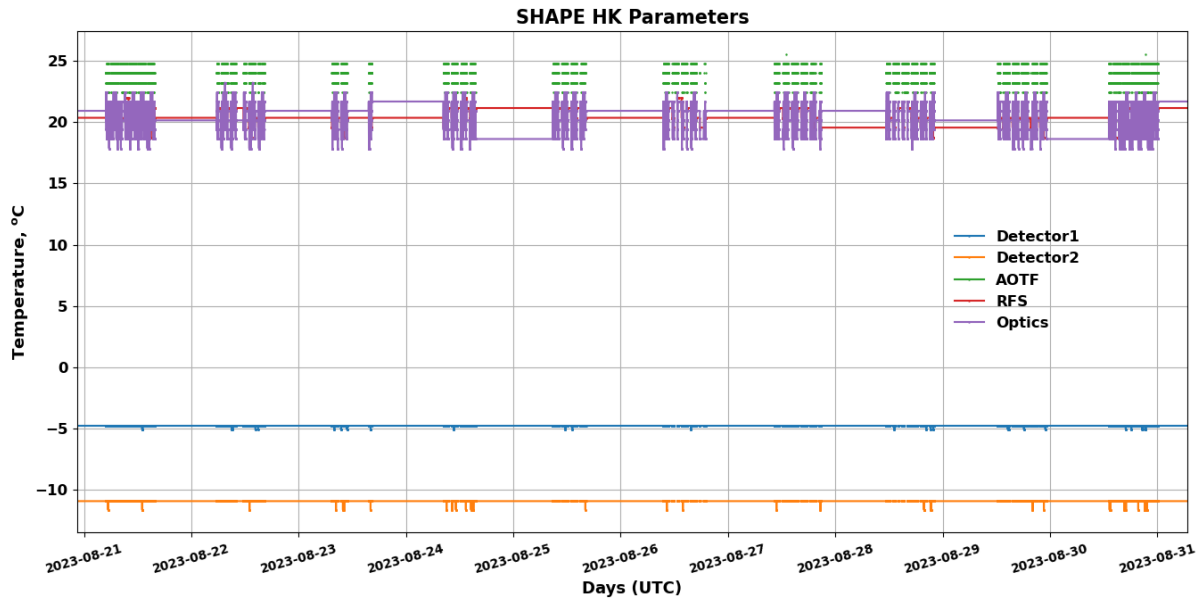

*Fig. 8: Variation of HK parameters (temperature of Detectors, AOTF, Optics, RF Power Amplifier) during the initial phase of in-flight performance verification.*

Fig. 8 shows the variations of a few HK parameters (Temperature of Detectors, AOTF, Optics, RF Power Amplifier) during payload operation over 10 days' time of initial observations till regularization of the operations. The efficient thermal management ensured tight control over all the temperature values of different sub-systems.

After SHAPE had been powered on, the alignment of the payload performance had to be verified in order to ensure that the alignment was intact post-launch. Fig. 9 shows the distribution of the spot size over the pixels in both detectors. The plot shows that the alignment was almost intact including the pointing uncertainties.

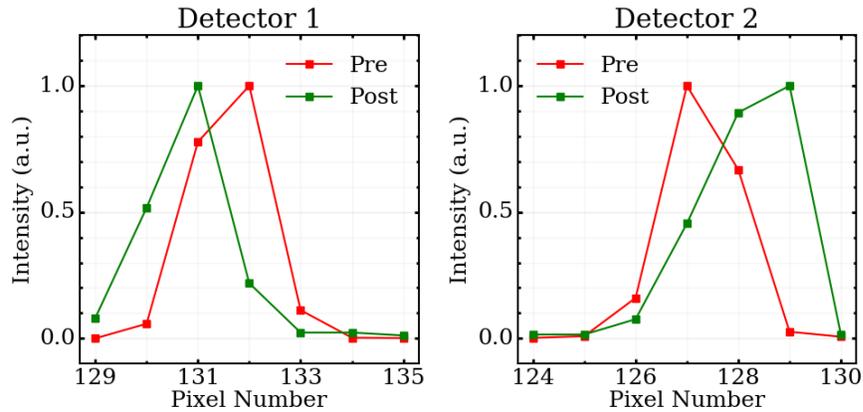

*Fig. 9: SHAPE spot (pixel distribution) pre- and post-launch (August 21, 2023) for both detectors. The intensity is normalized with respect to the maximum value to indicate the intensity distribution across the pixels.*

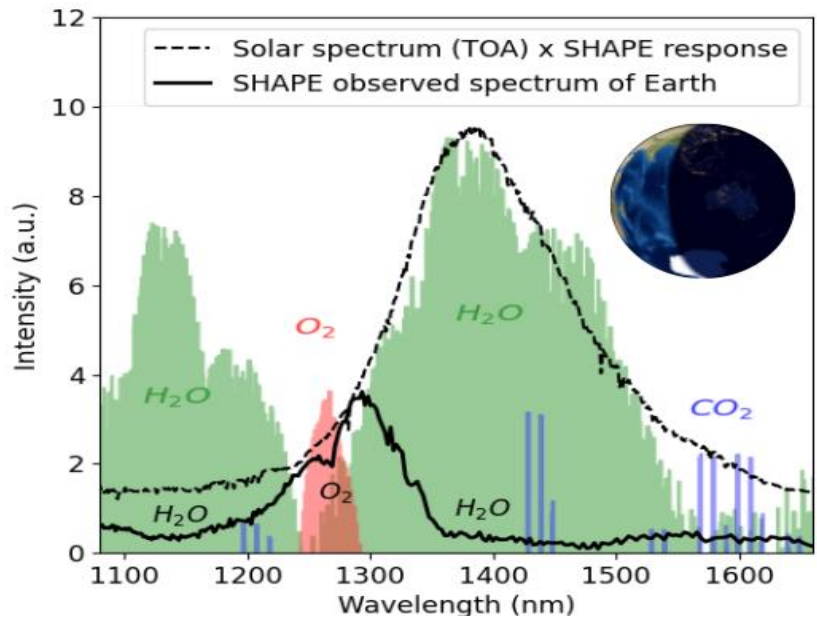

*Fig. 10: The Earth spectrum (black solid line) as recorded by SHAPE on August 26, 2023 is shown along with the solar spectrum (black dashed line) at the top of the atmosphere (TOA) as would be seen by SHAPE. The absorption line strengths of various atmospheric gases are shown in the background as colored lines (scaled-up for representation). The illuminated Earth at a phase angle $\Phi \sim 130^o$ as seen by SHAPE during this observation is shown in the inset.*

Since August 21, 2023, SHAPE has been carrying out Earth observations. Fig. 10 shows the spectrum recorded by Detector-1, where the absorption features corresponding to water and oxygen were detected. During this observation the Earth Phase angle was $\Phi \sim 130^O$ and the inset in the figure shows the view of Earth as seen from the Lunar orbit. SHAPE continues to study the Earth as an 'exoplanet' at different phase angles and generate data which will be used for further scientific analyses.

## Summary

The SHAPE is a unique 'first-of-its-kind' instrument to observe Earth as an 'exoplanet' from the lunar orbit. The instrument was realized within a quick turnaround time of 30 months. In this manuscript, we have highlighted the design aspects of the payload along with the overall functionality of the instrument. This includes details about the extensive testing as well as the calibration aspects of the instrument. After the successful separation of the Lander Module from the Propulsion Module, the payload has been in operation since August 21, 2023. Since then SHAPE has been operated in around 75 orbits for Earth observation as well as on-board calibration operations. We also presented the initial performance of the payload from the in-flight observations. At present, SHAPE is observing Earth on visibility basis and the science data collected from the payload will provide promising insights which will be vital for future 'exoplanet' missions.

## Acknowledgement

Authors thank the Group Head, Space Astronomy Group (SAG); Deputy Director, Payload Data Mgmt. & Space Astronomy Area; Associate Director and Director of U. R. Rao Satellite Centre (URSC) for encouragement and continuous support to carry out this work. Authors also thank Director, LEOS for constant support towards realization of the payload. The Indian Space Research Organization (ISRO) funded, managed and facilitated the overall project.